# Facile One Pot Synthesis of Hybrid Core-Shell Silica-Based Sensors for Live Imaging of Dissolved Oxygen and Hypoxia Mapping in 3D cell models

Helena Iuele,[1*] Stefania Forciniti,[1] Valentina Onesto,[1] Francesco Colella,[1,2] Anna Chiara Siciliano,[1,2] Anil Chandra,[3] Concetta Nobile[1], Giuseppe Gigli,[1,4] Loretta L. del Mercato[1*]

[1] Institute of Nanotechnology, National Research Council (CNR-NANOTEC), c/o Campus Ecotekne, via Monteroni, 73100, Lecce, Italy.

[2] Department of Mathematics and Physics "Ennio De Giorgi", University of Salento, c/o Campus Ecotekne, via Monteroni, 73100, Lecce, Italy

[3] Centre for Research in Pure and Applied Sciences, Jain (Deemed-to-be-university), Bangalore, Karnataka 560078, India

[4] Department of Experimental Medicine, University of Salento, c/o Campus Ecotekne, via Monteroni, 73100, Lecce, Italy

[*]: corresponding authors

[*]*E-mail addresses:* helena.iuele@nanotec.cnr.it (Helena Iuele); loretta.delmercato@nanotec.cnr.it (Loretta L. del Mercato)

**Key words:** Optical ratiometric oxygen-sensors; core-shell silica microparticles; alginate microgels; oxygen spatiotemporal tracking; cell metabolism; computational analysis.

## ABSTRACT

Fluorescence imaging allows for non-invasively visualizing and measuring key physiological parameters like pH and dissolved oxygen. In our work, we created two ratiometric fluorescent microsensors designed for accurately tracking dissolved oxygen levels in 3D cell cultures. We developed a simple and cost-effective method to produce hybrid core-shell silica microparticles that are biocompatible and versatile. These sensors incorporate oxygen-sensitive probes (Ru(dpp) or PtOEP) and reference dyes (RBITC or A647 NHS-Ester). SEM analysis confirmed efficient loading and distribution of the sensing dye on the outer shell. Fluorimetric and CLSM tests demonstrated the sensors' reversibility and high sensitivity to oxygen, even when integrated into 3D scaffolds. Aging and bleaching experiments validated the stability of our hybrid core-shell silica microsensors for 3D monitoring. The Ru(dpp)-RBITC microparticles showed the most promising performance, especially in a pancreatic cancer model using alginate microgels. By employing computational segmentation, we generated 3D oxygen maps during live cell imaging, revealing oxygen gradients in the extracellular matrix and indicating a significant decrease in oxygen levels characteristic of solid tumors. Notably, after 12 hours, the oxygen concentration dropped to a hypoxic level of $PO_2$ 2.7 ± 0.1%.

## 1.0 INTRODUCTION

Oxygen is an indispensable for cellular functions, playing a pivotal role in energy production and cellular signalling. It acts as the final electron acceptor in mitochondrial respiration and is a critical substrate for various enzymatic reactions involved in metabolism of carbohydrates, lipids, and other cellular components.[1] Additionally, oxygen serves as a key regulator of cellular signalling pathways. Indeed, dysregulation of oxygen homeostasis is implicated in various diseases like cancer, cardiovascular disorders, and neurodegenerative conditions.[2,3] Solid tumor microenvironments are characterized by hypoxia, nutrient depletion, increased stiffness, and acidosis due to poor vasculature.[4,5] Understanding the molecular mechanisms of the hypoxic microenvironment is crucial for the development of new targeted therapeutic protocols. Over the years, a variety of analytical methods have been developed for quantifying oxygen levels in biological systems, ranging from direct to indirect measurements of oxygen-sensitive parameters. Direct methods involve sensing dissolved oxygen directly in the sample and include Clark-type electrodes[6] and metabolic analyzers like the Seahorse Extracellular Flux Analyzer (Seahorse XF Analyzer, Agilent).[7] However, these approaches only provide mediated oxygen concentration measurements, and their bulk analysis does not assess the heterogeneity of oxygen distribution in tissues or 3D cell systems. Moreover, electrodes have limited applications due to their physical invasiveness,[8] while the Seahorse XF Analyzer necessitates dedicated instruments and kits, making its use relatively niche due to cost considerations. Indirect methods encompass imaging techniques like magnetic resonance imaging (MRI)[9] and mitochondrial NADH/$NAD^+$ fluorescence imaging,[10-13] which enable measurement of oxygen distribution at the cellular scale. However, these approaches are indirect analyses; they do not directly measure oxygen itself but instead assess oxygen-dependent parameters that provide insights into oxygen consumption in living specimens.

Nanotechnology has introduced novel tools for the direct, real-time measurement of $PO_2$ with micron-scale spatial resolution. They offer advantages like non-invasiveness, enhanced sensitivity for monitoring of oxygen distribution within tissues and *in vitro* systems[14] and include micro/nano optical sensors that exploit phosphorescent or fluorescent oxygen-sensitive probes to directly quantify oxygen levels in biological samples.[4] In this study, we optimized a versatile synthetic approach for producing optical ratiometric microsensors. The straightforward chemistry and sensor architecture ensure easy customization to fit different image settings, enabling automatic and accurate 3D oxygen tracking at the cellular level, even with less advanced imaging setups. These optical sensors are ratiometric, allowing their use with confocal or FLIM/PLIM technology. While fluorescence lifetime imaging offers greater accuracy than emission intensity analyses, it has drawbacks such as cost and the need for dedicated units,[15] moreover high power pulsed laser sources required by FLIM/PLIM can induce photobleaching and phototoxicity.[16] Additionally, image acquisition and analysis can be time-consuming,[17] making it less convenient for 3D time-lapse live imaging compared to the ratiometric approach, which shows to be more practical.[18] Lastly, our synthetic approach facilitates an easy scaling up, as the process is conducted in the same reactor at room temperature, using only ethanol and water as main solvents. Silica was selected as the core material for its inertness, biocompatibility, and the versatility to tailor its morphology and size from the nano- to the microscale. Utilizing a modified Stöber method,[19-21] we produced microparticles for the assembly of oxygen sensors, with the added flexibility of successful nanoparticle production. Opting for microparticles facilitated tracking extracellular oxygen gradients, allowing precise visualization and automated image segmentation during confocal imaging[22, 23]. The silica particles were additionally labelled with a reference dye such as Rhodamine B isothiocyanate (RBITC) or Alexa Fluor™ 647 NHS Ester (Alexa647), meanwhile, oxygen-sensitive complexes were encapsulated within an outer shell of oxygen-permeable poly(dimethylsiloxane) hydroxy terminated (PDMS-OH). In particular, we selected tris(4,7-diphenyl-1,10-phenanthroline)ruthenium(II) dichloride (Ru(dpp)) and Platinum(II) octaethylporphyrin (PtOEP) as sensing dyes because Ruthenium(II) complexes were extensively studied due to their optical performances,[24-26] customizable structures and reliability[27, 28] and Platinum (II) complexes were extensively used due to their suitability not only for *in vitro*[4] but also for *in vivo*[29] [30] live-cell imaging of both intracellular[31] and extracellular oxygen gradients.[32] Ru(dpp) and PtOEP dyes exhibit quenching of their fluorescence/phosphorescence upon interaction with molecular oxygen or reactive oxygen species,[33, 34] a characteristic shared with other metal-organic complexes of transition metals and metalloporphyrins.[6, 35-39] The incorporation of an outer PDMS-OH layer resulted effective in overcoming solubility limitations of the sensing dyes as well as avoiding their leaching, while guaranteeing oxygen permeability. Metal-organic complexes are often too hydrophilic[40] to cross cellular membranes, restricting their application to the extracellular environment, while some metalloporphyrins are too hydrophobic[41, 42] to be solubilized in aqueous media. This strategy not only adjusted the solubility of oxygen probes but also optimized their loading capacity, offering the potential to encapsulate other relevant oxygen sensing dyes such as dichlorotris(1,10-phenanthroline)ruthenium(II) chloride (Ru(Phen)) or tris(2,2'-bipyridyl)dichlororuthenium(II) hexahydrate (Ru(bpy)).[43]

In this study, we synthesized, morphologically characterized, calibrated using spectrofluorimetry and confocal laser scanning microscopy (CLSM), and *in vitro* validated two oxygen-sensitive ratiometric sensors. Our aim was to demonstrate the versatility of the developed synthetic approach, particularly focusing on assembling sensors using the Ru(dpp)-RBITC or PtOEP-Alexa647 systems as oxygen-sensitive and reference dyes, respectively. Among these, the Ru(dpp)-RBITC-based microsensors emerged as the most reliable system, proving to be the optimal systems for biological studies. To evaluate its performance, we conducted tests within an *in vitro* three-dimensional cancer model employing sodium alginate as a hydrogel scaffold. Pancreatic ductal adenocarcinoma (PDAC) was selected as the tumor model due to its characteristic severe hypoxia, with oxygen levels as low as 0.7%.[44, 45] Our investigations enabled the creation of spatial-temporal profiles for dissolved oxygen levels within the engineered tumor constructs, thereby confirming the applicability of our methodology for non-invasive recording of oxygen dynamics within tumor microenvironments and pharmacological evaluations.

## 2.0 EXPERIMENTAL SECTION

### 2.1 Chemicals and reagents

Rhodamine B isothiocyanate (RBITC, mW: 536.08 g/mol, Sigma Aldrich), platinum octaethylporphyrin (PtOEP, mW: 727.84 g/mol, Sigma Aldrich), ammonium hydroxide 28% (mW: 35.05 g/mol, Sigma Aldrich), hydrogen peroxide solution 30 wt% in $H_2O$ (mW: 34.01 g/mol,

Sigma Aldrich), tritonTM X-100 (Sigma Aldrich), alexa fluor™ 647 succinimidyl ester (A647 NHS-Ester, mW: 1025.2 g/mol, Thermo Fisher Scientific), tris(4,7-diphenyl-1,10-phenanthroline)ruthenium(II) dichloride (Ru(dpp), mW: 1169.17 g/mol, Alfa Aesar), tetraethyl orthosilicate (TEOS, mW: 208.33 g/mol, Sigma Aldrich), poly(Dimethylsiloxane) hydroxy terminated (PDMS-OH, mW: ~550 g/mol, Aldrich Chemistry), (3-Aminopropyl)triethoxysilane (APTES, mW: 221.37 g/mol, Sigma Aldrich), anhydrous ethanol (mW: 46.07 g/mol, VWR), Ethanol (mW: 46.07 g/mol, Honeywell), potassium chloride (KCl, mW: 74.55 g/mol, Sigma Aldrich), sodium chloride (NaCl, mW: 58.44 g/mol, Sigma Aldrich) Sodium hydroxide (NaOH, mW: 40.00 g/mol, Sigma Aldrich), Hydrochloric acid (HCl, mW: 36,46 g/mol Sigma Aldrich), Alginic acid sodium salt from brown algae (Sodium Alginate) (LOW EEO, Sigma life science), Calcium chloride ($CaCl_2$, mW: 110.98 g/mol, Sigma Aldrich).

## 2.2 Ru(dpp)-RBITC microsensor synthesis

Ru(dpp)-RBITC microsensor synthesis is represented in **Scheme 1a**. Silica seed particles were synthesized by dissolving 2.8 mg of KCl in 1.12 mL of milliQ water. 12 mL of ethanol, 760 μL of $NH_4OH$ 28% and 200 μL of TEOS were added to the mixture and the reaction was kept under stirring at 240 rpm, for 20 minutes, at room temperature (RT).

RBITC-APTES thiourea (with RBITC: APTES in 1:3 molar ratio) was performed by dissolving 0.2 mg of RBITC in 2 mL of anhydrous ethanol. 0.37 μL of APTES were added to the solution, and the reaction mixture was kept under stirring for 2 hours at RT. Then, 2 mL of anhydrous ethanol and 0.50 μL of TEOS were added to thiourea crude-reaction solution. This solution was filled into a 6.0 mL syringe and injected into the flask containing the silica seed particles with a flow rate of 0.05 mL/minute. The reaction mixture was kept under stirring at 240 rpm, for 2 hours at RT. A second silica shell was prepared by dissolving 0.59 mg of Ru(dpp) into 3.3 mL of anhydrous ethanol maintaining a 1:1 molar ratio between Ru(dpp) and the RBITC of the inner shell. Then 336 μL of TEOS and 336 μL of PDMS-OH were added to the solution and the mixture was filled into a 6 mL syringe and injected into the flask containing the silica particles, with a flow rate of 0.05 mL/minute. The reaction was left under stirring at 240 rpm for 20 hours RT. The final product was centrifuged at 1500 rpm for 5 minutes and the supernatant was discarded. The pellet was washed in 40 mL of ethanol three times (1500 rpm, 5 minutes) and then in distilled water. The clean product was resuspended in ethanol in order to obtain a 40 mg/mL stock solution and stored in the dark at room temperature.

*2.3 PtOEP-Alexa647 microsensor synthesis*

PtOEP-Alexa647 microsensor synthesis is represented in **Scheme 1b**. Silica seed particles were synthesized by dissolving 2.8 mg of KCl in 1.12 mL of milliQ water. 12mL of ethanol, 760 μL of NH4OH 28% and 200 μL of TEOS were added to the reaction mixture, which was kept under stirring at 240 rpm, for 20 minutes, at RT.

Alexa647-APTES thiourea synthesis (with Alexa647: APTES in 1:3 molar ratio) was performed by dissolving 0.5 mg of Alexa647 in 100 μL of DMSO. 2 mL anhydrous ethanol and 0.37 μL of APTES were added to such solution and the reaction was kept under stirring for 2 hours at RT. Subsequently, 2 mL of anhydrous ethanol and 0.50 μL of TEOS were added to the thiourea crude-reaction product and this solution was filled into a 6 mL syringe and injected into the flask containing the silica-seed particles, with a flow rate of 0.05 mL/minute. The reaction mixture was kept under stirring at 240 rpm, for 2 hours at RT.

The second silica shell was prepared by dissolving 1.73 mg of PtOEP in 500 μL of chloroform and by adding this solution into 3.3 mL of anhydrous ethanol (considering a 1:6 molar ratio between Alexa647 and PtOEP, respectively. Then, 336 μL of TEOS and 336 μL of PDMS-OH were added to the solution. This mixture was filled into a 6 mL syringe and injected into the flask containing the silica particles with a flow rate of 0.05 mL/minute. The reaction was carried under stirring at 240 rpm for 20 hours at RT.

The final product was centrifuged at 1500 rpm for 5 minutes and the supernatant was discarded. The pellet was washed in 40 mL ethanol three times (1500 rpm, 5 minutes) and then in distilled water other 3 times (1500 rpm, 5 minutes). The clean product was resuspended in ethanol in order to obtain a 40 mg/mL stock solution and stored in the dark at room temperature.

*2.4 Dynamic light scattering*

The size, dimension, monodispersity and surface charge of Ru(dpp)-RBITC and PtOEP-Alexa647 sensors were monitored during the synthesis and the washing steps via Dynamic Light Scattering (DLS) (Zetasizer Nano ZS, MALVERN) analysis (refractive index in water 1.458, absorption 0.010, 25°C, 2 min equilibrium time[46]). Data were collected and elaborated using the Zetasizer family software v7.12.

## 2.5 Flow cytometry analysis

Ru(dpp)-RBITC or PtOEP-Alexa647 sensors were prepared at the concentration of 0.02 mg/mL in PBS 1X containing 5% FBS and 0.5% Triton X-100 in order to avoid their aggregation. After 30 minutes of incubation in a sonicator bath (Digital ultrasonic cleaner, Argo Lab), microparticles were carefully pipetted up and down to reduce the aggregates and then analysed on a flow cytometer (CytoFLEX S, Beckman Coulter) with high resolution for nano- and microparticles measurement reached by using the side scatter of the 405 nm laser. The number of single events contained in 10 µL of solution was acquired and analysed using CytExpert software. Debris were excluded based upon forward scatter and side scatter measurements.

## 2.6 Electron microscopy measurements

The morphological characterization of Ru(dpp)-RBITC and PtOEP-Alexa647 sensors has been carried out adopting a transmission electron microscope (TEM, JEM 1400Plus, JEOL Ltd., Japan), equipped with LaB6 filament-source, operating at 120 kV.. 10 µL of a 4 mg/mL Ru(dpp)-RBITC/PtOEP-Alexa647 ethanoic solution were drop-casted onto standard carbon-coated copper TEM grids, allowing the solvent to evaporate. The sample grids were further dried under vacuum overnight prior being transferred into the microscope for imaging. .

The same samples, prepared on carbon-coated copper TEM grids, after being transferred onto dedicated SEM holders, were then analyzed adopting a scanning electron microscope (SEM, Carl Zeiss Merlin FE-SEM microscope, Germany) equipped with a Gemini II column, with a FEG source, operating at 20 kV, with a few seconds of acquisition time. SEM images of Ru(dpp)-RBITC and PtOEP-Alexa647 were finally analysed via ImageJ [47] software in order to measure sensor diameters and distributions.

## 2.7 Static contact angle measurements

The hydrophobicity of Ru(dpp)-RBITC and PtOEP-Alexa647 was assessed by measuring their static water contact angles (WCA) adopting a CAM 200 (KSV Instruments Ltd., Finland) instrument. A thin and homogeneous layer of sensors stock solution (~ 100 µL) was deposited on the surface of a cover glass (24 X 24 mm, 0.13-0.17 mm thickness) (**Figure S1**). The samples were left drying overnight at room temperature. For the experiment set up, the liquid (heavy phase) was represented by water (d= 0,9986 g/cm$^3$) and the liquid (light phase) was air (d=0,0013 g/cm$^3$). The measurements were performed by loading on the tip of the needle a drop of 2 μL of volume that was stroke on the surface of the sample.

### 2.8 Calibration, reversibility, aging and dye leaching evaluation

2 μL of Ru(dpp)-RBITC or PtOEP-Alexa647 sensors stock solution were diluted in 250 μL of Leibovitz's L-15 medium no phenol red (L-15, Gibco) for calibration and for reversibility, after having dispersed the stock solutions by vortexing.

Sensors were calibrated starting from 0.5% up to atmospheric PO$_2$ of 19.6% adopting a CLARIOstar® Plus plate reader (CLARIOstar, BMG LABTECH Inc., USA) equipped with an Atmospheric Control Unit (ACU) while data were acquired and analyzed using MARS Data Analysis software provided by the manufacturer. The ACU was purging N$_2$ into a multiwell chamber containing Ru(dpp)-RBITC or PtOEP-Alexa647 sensors dissolved in L-15 medium while O$_2$ concentration was continuously monitored. Five acquisitions for each known PO$_2$ (0.5%, 5.0%, 10.0%, 15.0%, 19.6%) were recorded and the calibration curve was extracted.

The aging of the sensors was evaluated by storing them at 4 °C in L-15 medium and the same calibration was repeated after 10 days.

The reversibility was evaluated by recording three cycles of switches between atmospheric PO$_2$ of 19.6 % and 0.5 % and the measurements were repeated after 10 days to evaluate their aging. Ru(dpp)-RBITC sensors were excited at $\lambda_{ex}$ = 468 nm and $\lambda_{ex}$ = 561 nm, fluorescence emission signals were collected in the 500-700 nm range and 570-700 nm range, respectively. PtOEP-Alexa647 sensors were excited at $\lambda_{ex}$ = 542 nm and $\lambda_{ex}$ = 650 nm, fluorescence emission signals were acquired in the 570-700 nm range and 670-700 nm range, respectively.

Possible dye leaking was evaluated after 1, 3 and 10 days of aging by diluting 120 μL of samples in 1500 μL of L-15 medium. At the selected times, sensors were centrifuged (6000 rpm, 45 seconds) and surnatant was sieved using a 1 μm pluriStrainer® filters and the absorbance was recorded in the range 400-800 nm (1 nm resolution).

### 2.9 Evaluation of sensors selectivity

Ru(dpp)-RBITC or PtOEP-Alexa647 sensors at fixed PO$_2$ of 19.6 % were tested at various pH values, in the 5.0-8.0 range, at different concentrations of Na$^+$ or K$^+$, in the range 0-160 mM and at different concentrations of H$_2$O$_2$, in the range 0-30 mM. 100 μL of stock solution were added to 20 mL of L-15 media, then the pH was adjusted with NaOH 1M or HCl 1M and monitored by a pHmeter (XS50, XS Instruments, Italy) at selected pH values (5.0, 5.5., 6.0, 6.5, 7.0, 7.5, 8.0). Similarly, 6 µL of sensors stock solution were added to 1 mL of distilled water at selected $Na^+$ or $K^+$ dilutions in the physiological range (0, 20, 40, 60, 80, 100, 120, 140, 160 mM) while 6 µL of sensors stock solution were added to 1 mL of L-15 media adjusted at physiologically relevant $H_2O_2$ concentrations (0, 5, 10, 20, 30 mM). Finally, aliquots of each sample were analysed at the plate reader by adopting the same setting mentioned in section 2.8.

## 2.10 Biocompatibility of sensors

The biocompatibility of Ru(dpp)-RBITC and PtOEP-Alexa647 sensors was evaluated by *in vitro* cytotoxicity assays using human cervical carcinoma cells (HeLa), human pancreatic cancer cells (AsPC-1), breast cancer cells (MCF7) and 3T3 fibroblast cell line. Cells were obtained from American Type Culture Collection (ATCC, Rockville, Md., USA) and cultured at 37 °C in a humidified 5% $CO_2$. AsPC-1 were routinely cultured in RPMI-1640 (Sigma-Merck KGaA, Darmstadt, Germany) whereas HeLa, MCF7 and 3T3 were grown in DMEM medium (Sigma-Merck KGaA, Darmstadt, Germany), both supplemented with 10% Fetal bovine Serum (FBS, Gibco), 2mM L-glutammine and 100 U/mL penicillin and streptomycin (Sigma-Merck KGaA, Darmstadt, Germany). The cell lines ($50 \times 10^4$ cells/well) were seeded in 24-well plates and their viability was evaluated through MTT assay (Sigma-Aldrich, Darmstadt, Germany). Cells were treated with Ru(dpp)-RBITC or PtOEP-Alexa647 sensors at the final concentration of 0.4 µg/mL starting from Ru(dpp)-RBITC stock solutions of 35.6 mg/mL and PtOEP-Alexa647 stock solutions of 40 mg/mL. Untreated cells were used as control. After 24 and 48 hours of treatment, 50 µL of MTT solution (0.5 mg/mL final concentration) were added into each well and incubated for 1 h at 37°C and 5% $CO_2$. In order to dissolve the formazan crystals, acidified isopropanol (2M HCl in isopropanol) was added to each well (1:1 ratio) and properly mixed. The absorbance was measured at 590 nm using a microplate reader (ClarioStarPlus, BMG Labtech). The medium without cells and containing Ru(dpp)-RBITC or PtOEP-Alexa647 sensors was used as blank subtracted from the absorbance values of the samples.

## 2.11 Production of three-dimensional (3D) oxygen sensing alginate scaffolds

The 3D oxygen sensing scaffolds were produced by adapting the methodology previously described by Rizzo et al.[48] (**Figure S2**). Briefly, human pancreatic tumour cells, AsPC-1 (ATCC CRL-1682) were cultured at 37 °C in a humidified 5% $CO_2$ incubator according to ATCC protocols. Human immortalized pancreatic stellate cells (PSCs), kindly provided by Dr. Enza Lonardo (Institute of Genetics and Biophysics of Cnr, Naples, Italy) were cultured in DMEM medium (Sigma-Merck KGaA, Darmstadt, Germany) supplemented with 10% FBS (Gibco), 2 mM glutamine and 1% penicillin/streptomycin (Sigma-Merck KGaA, Darmstadt, Germany) at 37 °C with 5% $CO_2$. Tumor cells, prelabelled with CellTracker™ Deep Red (Invitrogen, ThermoFisher Scientific) and PSC cells were mixed in a 1:3 ratio in 250 µL of culture medium.[49] Then, 250 µL of 3% (w/v) alginate (dissolved in deionized and sterile water) were added to the cell suspension with 40 µL of oxygen sensors and gentle mixed for 5 minutes. Subsequently, the alginate solution, containing cells and sensors was collected into a syringe (BD Plastipak™ 1-mL Syringe) and placed into a syringe pump. By applying a high-voltage generator, alginate microgels were formed once the suspension droplets felt into the calcium chloride solution (100 mM $CaCl_2$ and 0.4% w/v of Tween-20 dissolved in water). Next, the alginate microgels embedding cells and oxygen sensors, were washed with fresh culture medium to remove the unreacted calcium chloride solution, and kept in the incubator under controlled temperature and 5% $CO_2$ before been processed for CLSM live imaging. In parallel, alginate microgels embedding oxygen sensors without cells were used for the calibration of the system through their exposure to known oxygen concentrations as described in section 2.12.

## 2.12 Confocal laser scanning microscopy (CLSM) characterization

Ru(dpp)-RBITC and PtOEP-Alexa647 morphological characterization was performed via super-resolution (LSM980 microscope equipped with Airyscan confocal super-resolution, Zeiss). Images were acquired with a 63x oil immersion objective, adopting Airyscan in resolution mode and employing 2.5 (resolution 1267x1267; field of view 53.87 x 53.57 µm) or a 5.0 as zoom (resolution 1024x1024; field of view 26.94 x 26.94 µm). During the imaging, Ru(dpp) was excited with $\lambda_{ex}$=488nm ($\lambda_{em}$=400-750 nm), RBITC and PtOEP were excited with $\lambda_{ex}$=543nm ($\lambda_{em}$=400-750nm) and Alexa 647 was excited with $\lambda_{ex}$=639 nm ($\lambda_{em}$=400-750 nm).

For CLSM analyses, alginate microgels or 2 µL of the stock solution of free sensors added to 300 µL of L-15 were deposited into a µ-Slide 8 Well Chamber Slide (IBIDI) previously treated with 0.3 mg/mL of Poly-L-lysine (37 °C, 30 min). Calibrations and live imaging were performed via CLSM (LSM 700 microscope, Zeiss) equipped with Okolab Stage Top Incubator and LEO hand-held meter (Okolab s.r.l., Italy), which was equilibrated to 37°C, 95% relative humidity and $PO_2$ in the 1-18 % range. The calibrations of the sensors were acquired with a 63x oil immersion objective and 2.5 as zoom (resolution 1024x1024; field of view 40.65 x 40.65 µm), while the calibrations and the timelapse experiments of alginate microgels were acquired with a 20x objective, 1.5x zoom and 2.55 µm step size (resolution 1024x1024; field of view in 213.39 × 213.39 µm; 29 sections). During CLSM measurements, Ru(dpp) was excited with $\lambda_{ex}$=488nm ($\lambda_{em}$=400-700 nm), RBITC and PtOEP were excited with $\lambda_{ex}$=555nm ($\lambda_{em}$=570-700nm), AsPC-1 stained with Deep Red and Alexa 647 were excited with $\lambda_{ex}$=639 nm ($\lambda_{em}$=640-700 nm) keeping the pinhole at 1 AU.

Gain and offset of the different laser lines were optimized at $PO_2$ 1%. Prior timelapse imaging, the oxygen content of the fresh L-15 media at time 0 (as control) was also measured adopting an optical probe (Vernier Go Direct® Optical dissolved oxygen probe).

### 2.13 Image segmentation and analysis

Confocal images of Ru(dpp)-RBITC and PtOEP-Alexa647 sensors were analysed with GNU Octave (version 7.2.0) with a custom algorithm[48, 50] to automatically extract the microparticle fluorescence intensity for ratiometric analyses. Briefly, 3D (*x,y,z*) images of the reference (RBITC or Alexa647) channels were first converted to grayscale and binarized with Otsu's method.[51] Then, objects (*i.e.*, the oxygen sensors) in 3D were identified by direct connectivity of the binary images along the z-axis, obtaining a 3D binary matrix, which was used as a mask to extract positions and mean fluorescence intensity ratios of the particles belonging to the original indicator and reference dyes images. Specifically, for each sensor particle, the pixel-by-pixel ratio between indicator and reference fluorescence intensities was calculated and averaged over the number pixels composing that sensor. Oxygen values were extracted by using a previously obtained calibration curve.

### 2.14 Statistical analysis

All experiments were performed in triplicate and the results were expressed as mean ± standard error. One-way analysis of variance (ANOVA) was performed to compare multiple conditions, and the Student's t-test was used for individual group comparison. Differences were considered significant with p-values < 0.05.

## 3.0 RESULTS AND DISCUSSION

*3.1 Synthesis of the oxygen sensing microparticles*

Ru(dpp)-RBITC and PtOEP-Alexa647 oxygen sensing microparticles were both obtained by a one-pot strategy based on a modified Stöber process.[52] This bottom-up approach, that exploits the hydrolysis of tetraethylorthosilicate for the sol-gel transition, is the most used for the production of silica nano- and microparticles with a diameter ranging from 50 to 2000 nm.[53] As shown in **Scheme 1a**, we produced silica core particles that were functionalized as ormosil (organically modified silica), with the thiourea obtained from the reaction between (3-aminopropyl)triethoxysilane and RBITC. These RBITC-labelled microparticles were then used for the non-covalent immobilization of oxygen sensing metal-organic complexes. Ru(dpp) was embedded in the microparticles by using both TEOS and PDMS-OH for the outer shell growth. The same approach was used for the synthesis of PtOEP-Alexa647 sensors (**Scheme 1b**) by employing Alexa 647 as reference dye. The obtained microsensors were then purified by centrifugations and resuspensions in ethanol, without any further purification steps. Both synthesis procedures are straightforward processes, carried out under mild conditions, requiring very few purification steps, making the whole procedure affordable and easy to be scaled up.

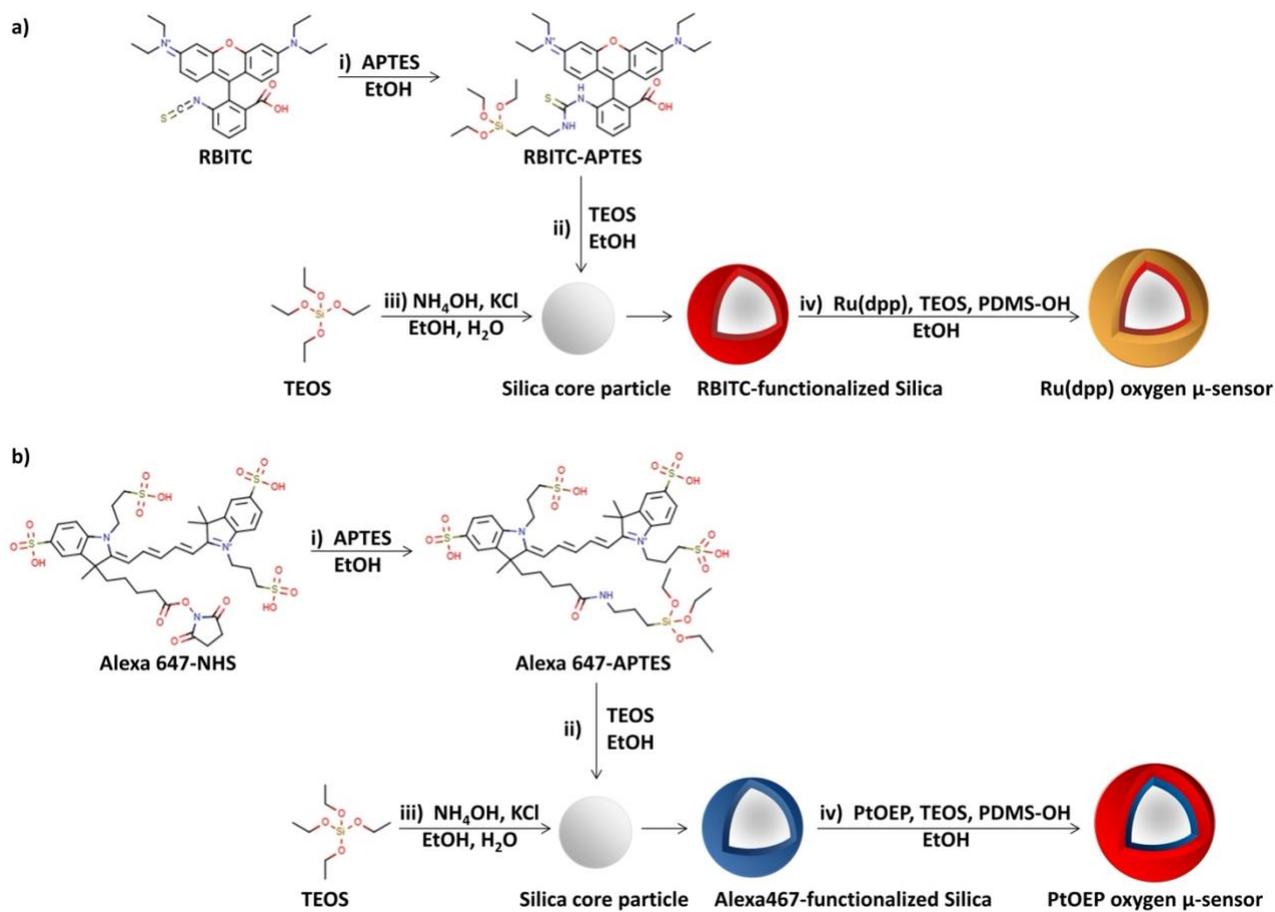

**Scheme 1. a)** Synthesis of Ru(dpp)-RBITC oxygen sensing microparticles: **(i)** RBITC reference dye coupling to APTES linker; **(ii)** silanization and regrowth of the silica core particles with RBITC-APTES thiourea and TEOS monomer, respectively; **(iii)** silica core particles formation in basic environment and in presence of KCl salt starting from TEOS as monomer; **iv)** deposition of the PDMS-OH shell containing Ru(dpp) oxygen sensing dye and regrowth of the silica core particles with TEOS on to the RBITC functionalized silica core particles. **b)** Alexa647 reference dye coupling to APTES linker; **(ii)** silanization and regrowth of the silica core particles with Alexa647-APTES amide and TEOS monomer, respectively; **(iii)** silica core particles growth in basic environment starting from TEOS monomer; **iv)** deposition of the PDMS-OH shell containing PtOEP oxygen sensing dye and regrowth of the silica core particles with TEOS on to the Alexa647 functionalized silica core particles.

## 3.2 Characterization of oxygen sensing microparticles

Size, charge, morphology and number of particles per volume of the oxygen-sensing microparticles were investigated via dynamic light scattering (DLS), electronic microscopy (SEM, TEM), static contact angle, Airyscan confocal super-resolution and flow cytometry. Ru(dpp)-RBITC oxygen sensors resulted to have an average hydrodynamic diameter of 1530 nm ± 55 nm (PdI 0.0013) (**Figure 1a**) and a zeta potential of -43.8 ± 2.1 mV, while PtOEP-Alexa647 oxygen sensors resulted to have and average hydrodynamic diameter of 1741 nm ± 186 nm (PdI 0.011) (**Figure 1b**) and a zeta potential of -48.1 ± 2.7 mV. Such difference in microparticle sizes might be related to the slightly different synthesis protocols, since the Stöber process is highly influenced by chemical and physical parameters such as solvents, type of base, potassium chloride concentration, temperature and reaction mixture viscosity[53]. Hydrodynamic size measurements of silica tagged with only RBITC or Alexa647 reference dyes were recorded and were compared with those of the fully assembled sensors to further assess the presence of the PDMS-OH polymer shell embedding Ru(dpp) dye or PtOEP dye. Notably, $SiO_2$-RBITC resulted to have an average hydrodynamic diameter of 1248 ± 73 nm (PdI 0.003) (**Figure 1a**), while $SiO_2$-Alexa647 presented a diameter of 1391 ± 80 nm (PdI 0.003) (**Figure 1 b**) showing a size approximately ≈ 300 nm smaller than the size of the oxygen sensing microparticles (1530 nm ± 55 nm and 1741 nm ± 186 nm, respectively) suggesting the presence of the PDMS-OH shell in both systems. SEM analyses, together with the average particle diameters, revealed morphological differences between plain and functionalized microparticles. For instance, silica microparticles (**Figures 1c-d**) exhibited a regular spherical shape and slight corrugations on the surface, due to silica porosity[52] and uniform diameter distribution with an average diameter of 1044 ± 23 nm (PdI 0.0005). Conversely, because of their outer shell, constituted by Ru(dpp) immobilized in PDMS-OH, Ru(dpp)-RBITC sensors (**Figures 1i-j**) displayed a rough surface and a higher average diameter of 1356 ± 62 nm (PdI 0.002).[54] The outer shell of PtOEP-Alexa647 sensors (**Figures 1o-p**) appeared less corrugated, more homogeneous and with a bigger average diameter of 1607 ± 64 nm (PdI 0.002), compared to the one of Ru(dpp)-RBITC sensors, despite a similar synthetic protocol. This could be ascribed to the different PtOEP polarity as well as to its different interaction with PDMS-OH.[35] In general, diameters extracted from SEM images were in accordance with the values reported by DLS analyses. Nevertheless, hydrodynamic sizes resulted being bigger than the ones evaluated in dry conditions, likely due to the dynamic swelling in water of the PDMS-OH layer.[54] To further evaluate the thickness of PDMS-OH polymer deposited on top of the silica cores, TEM analyses were performed. Silica microparticles (**Figures 1e-f-g**) exhibited a spherical shape with a smooth surface while Ru(dpp)-RBITC (**Figures 1k-l-m**) and PtOEP-Alexa647 (**Figures 1q-r-s**) showed a more corrugated surface characterized by the presence of a shell of approximately 40 nm and 10 nm, respectively confirming the presence of PDMS-OH polymer layer. Contact angle measurements showed that the silica microparticles (θ=53,43 ± 6,47) (**Figures 1h**) are hydrophilic with a static contact angle θ<90°, while Ru(dpp)-RBITC (θ=149,68±10,19) (**Figures 1n**) and PtOEP-Alexa647 sensors (θ=146,27 ± 4,04) (**Figures 1t**) are hydrophobic being θ >90°. These differences could be due to the presence of the hydrophobic PDMS-OH outer shell on the assembled Ru(dpp)-RBITC and PtOEP-Alexa647 sensors. Airyscan confocal super-resolution measurements were performed to localize the reference and sensing dyes and to evaluate their distribution in the assembled sensors. Ru(dpp)-RBITC imaging revealed the presence of two concentric rings: the inner ring corresponded to the RBITC while the outer ring to Ru(dpp) (**Figures 1u, S3**). Profile analyses performed on several Ru(dpp)-RBITC sensors (**Figures 1w-x**) revealed an average distance of 37 ± 9 nm between the RBITC and the Ru(dpp) maximum intensities. These distances are in accordance with Ru(dpp)-RBITC shell measurements via TEM analyses. The imaging of PtOEP-Alexa647 sensors also revealed the presence of two concentric rings with the inner one corresponding to Alexa647 and the outer one to PtOEP (**Figures 1v, S4**). Profile analyses performed on several PtOEP-Alexa647 sensors (**Figures 1w-x**) revealed an average distance of 15 ± 3 nm between the PtOEP and Alexa647, in accordance with TEM measurements. The number of microsensors ascertained via flow cytometry analyses resulted 1704.4 ± 59.6 μ-particles/μL (85220 μ-particles/mg) for Ru(dpp)-RBITC while 184.3 ± 33.4 μ-particles/μL (9210 μ-particles/mg) for PtOEP-Alexa647, accordingly with the different sensors dimension evaluated via SEM and DLS analyses.

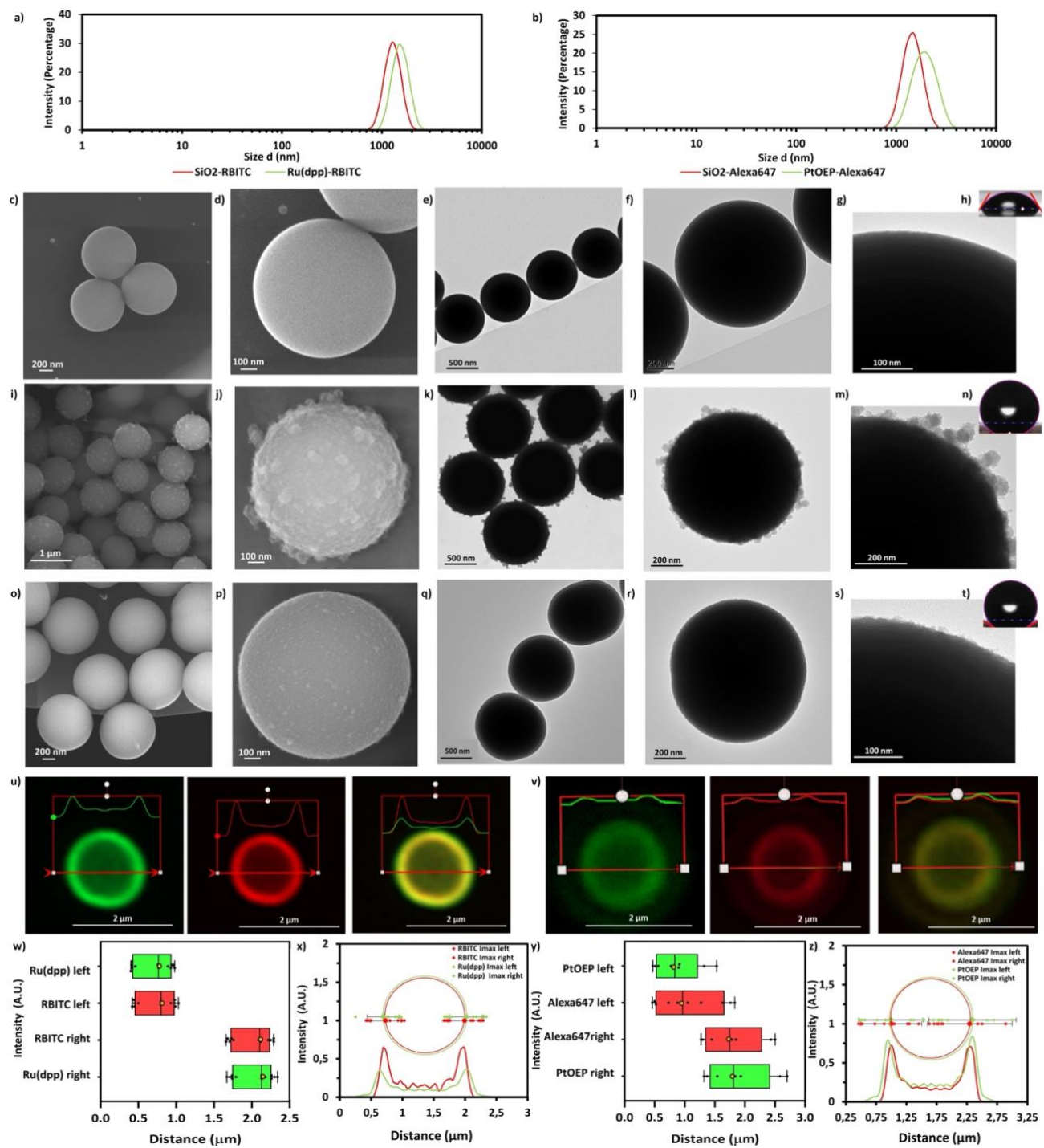

**Figure 1.** Representative DLS graphs of the hydrodynamic diameter distribution of a**)** SiO$_2$-RBITC particles and Ru(dpp)-RBITC sensors and **b)** SiO$_2$-Alexa647 particles compared to PtOEP-Alexa647 sensors. (refractive index in water 1.458; absorption 0.010; 25°C). Representative low and high magnification SEM image galleries of **c-d)** control silica particles, **i-j)** Ru(dpp)-RBITC sensors and **o-p)** PtOEP-Alexa647 sensors. Representative low and high magnification TEM image galleries and surface details of **e-f-g)** silica particles, **k-l-m)** Ru(dpp)-RBITC sensors and **q-r-s)** PtOEP-Alexa647 sensors. Static contact angle of **h)** control silica

particles, **n)** Ru(dpp)-RBITC sensors and **t)** PtOEP-Alexa647 sensors. Airyscan confocal superresolution microscopy of **u)** Ru(dpp)-RBITC sensors (green channel Ru(dpp): λex 488 nm, λem 400-750 nm; red channel RBITC: λex 543 nm, λem 400-750 nm and overlay of the fluorescence channels. Scale bars: 2 μm) **v)** PtOEP-Alexa647 sensors (false green channel PtOEP: λex 543 nm, λem 400-750 nm) and red channel (Alexa647: λex 639nm, λem 400-750 nm, and overlay of the fluorescence channels. Scale bars: 2 μm). **w)** Box Plot of maximum Intensity of Ru(dpp) green channel and RBITC red channel in function of the distance derived from **u)** intensity profile measurements. **u)** Average intensity profile of Ru(dpp) green channel and RBITC red channel in function of the distance. (Average Ru(dpp) and RBITC maximum Intensity in function of the distance interpolated with the corresponding circumference to evaluate Ru(dpp) shell thickness) **v)** Box Plot of maximum Intensity of PtOEP fake green channel and Alexa647 red channel in function of the distance derived from **v)** intensity profile measurements. **z)** Average intensity profile of PtOEP fake green channel and Alexa647 red channel in function of the distance. (Average PtOEP and Alexa647 maximum Intensity in function of the distance interpolated with the corresponding circumference to evaluate PtOEP shell thickness).

### 3.3 Ru(dpp)-RBITC and PtOEP-Alexa647 microsensors respond to rapid changes in oxygen and are stable over time

To assess whether Ru(dpp)-RBITC and PtOEP-Alexa647 oxygen microsensors are accurate, sensitive and stable over time, we performed calibration measurements and reversibility studies. Indeed, one of the main concern for these systems can be the reversibility and the rapidity of their response to oxygen concentrations; notably, these properties are intrinsic to the employed metal-organic dyes, and for this reason they are difficult to be tailored.[55] For calibration tests, oxygen microsensors were dispersed in L-15 media and $PO_2$ was adjusted at selected values (0.5 - 19.6 % $PO_2$ range). The calibration curves of Ru(dpp)-RBITC sensors and PtOEP-Alexa647 sensors obtained through spectrofluorometry analysis are reported in **Figures 2a-b** and are compared to those of the free dyes in solution (**Figure S5**) to evaluate if coupling and immobilization of the dyes to the final sensors could affect sensing behaviour. Ideally, the interaction between molecular oxygen and the employed metal-organic complexes should quench Ru(dpp) and PtOEP fluorescence signals in a proportional way, following the Stern-Volmer equation [56] (further details described in **SI**) . Indeed, the obtained results (**Figure 2a-b**) proved that the coupled and immobilized dyes in the assembled sensors maintain a similar behaviour to that of the free dyes in solution (**Figure S5**) and indicate that the chosen strategy of entrapment preserves the sensibility of the oxygen probes. Both microsensors are sensitive to dissolved $O_2$ concentration; notably, the fluorescence intensity ratio ($I_{Ru(dpp)}/I_{RBITC}$ or $I_{PtOEP}/I_{Alexa647}$) decreases as a function of oxygen concentration. The slope of the calibration curve is also a parameter useful to assess the sensors sensitivity towards oxygen. In our case Ru(dpp)-based sensors were more sensitive compared to the PtOEP ones (**Figures 2a-b**), due to the higher slope in the calibration curve (0.020 for Ru(dpp) and 0.013 for PtOEP). Sensor's reversibility and aging were also investigated (**Figure 2**). Reversibility was evaluated by a series of switches between $PO_2$ of 19.6 % and 0.5 % (**Figure 2c-d**). Ru(dpp)-RBITC sensors appeared more sensitive compared to PtOEP-Alexa647 which showed lower ratios values. As regards the sensors aging, calibrations (**Figure 2a-b**) and the same series of $PO_2$ switches (**Figure 2c-d**) were repeated after 10 days on the same samples for studying their sensing ability despite the aging. Although the aging led to a slight decrease in fluorescence signals, all observed differences resulted not statistically significant accordingly to ANOVA with single factor ($α=0.05$) thus proving that both the systems preserved their sensitivity. Finally, dyes leaking from the assembled sensors was evaluated over time. Absorbance measurements (**Figure S6a-b**) demonstrated high sensors robustness with negligible absorbance in the range 400-800 nm after 10 days of aging in L-15 media at 4°C.

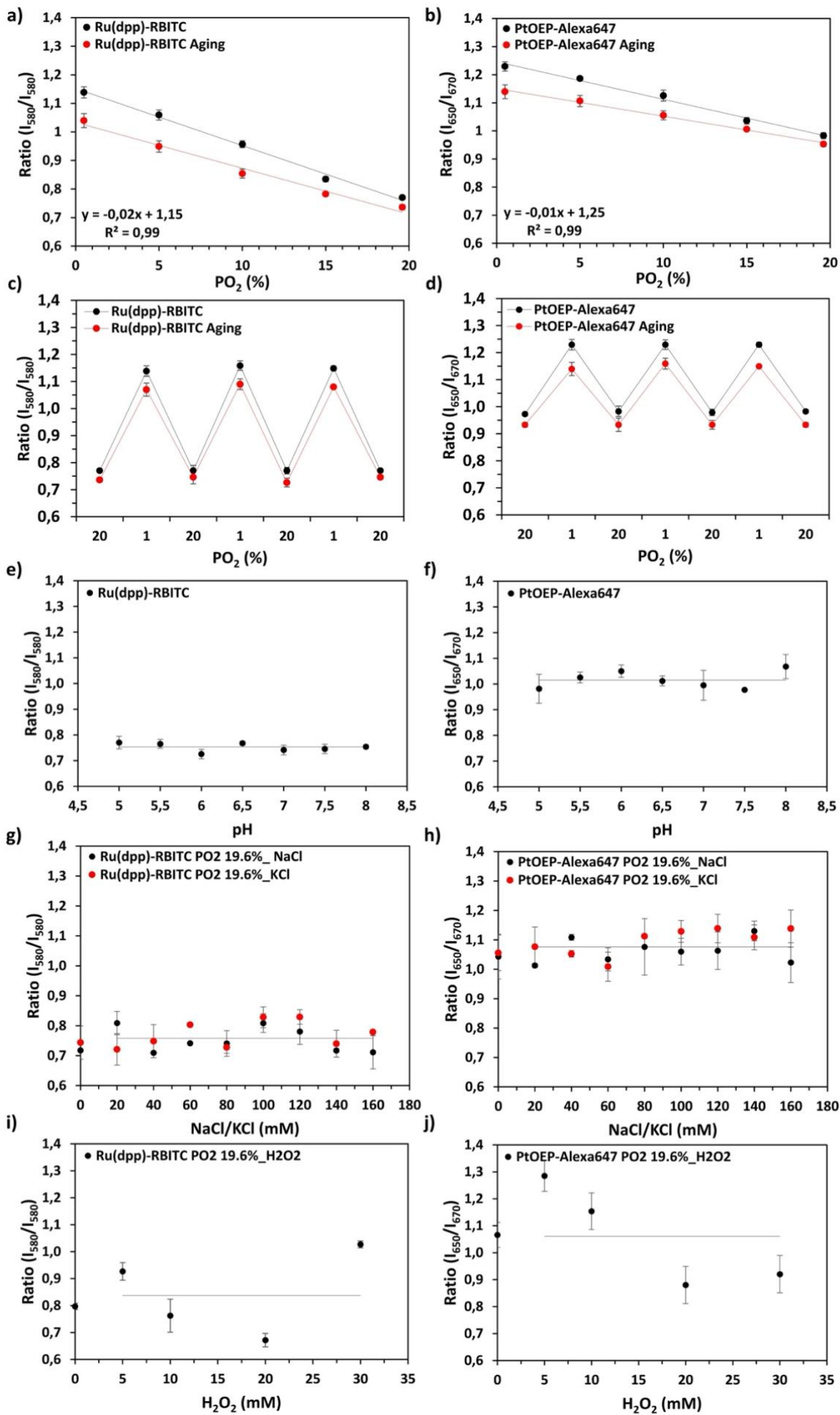

**Figure 2.** Oxygen-sensitive microsensors fluorimetric calibration, reversibility and selectivity: **a)** Ru(dpp)-RBITC ratios and **b)** PtOEP-Alexa647 ratios at time 0 and after 10 days of aging under increasing oxygen concentrations; **c)** Ru(dpp)-RBITC ratios and **d)** PtOEP-Alexa647 ratios (PO$_2$ 19.6 – 0.5, 3 cycles) at time 0 and after 10 days of aging. **e)** Ru(dpp)-RBITC ratios and **f)** PtOEP-Alexa647 ratios at different pH-adjusted media in the physiological range (5 - 8). **g)** Ru(dpp)-RBITC ratios and **h)** PtOEP-Alexa647 ratios in presence of Na$^+$/K$^+$ at physiological concentration range (0 – 160 mM). **i)** Ru(dpp)-RBITC and **j)** PtOEP-Alexa647 ratios in presence of H$_2$O$_2$ at physiological concentration range (0 – 30 mM). Ru(dpp)-RBITC and PtOEP-Alexa647 oxygen-sensitive microsensors emission spectra recorded with 1 nm of resolution, Ru(dpp): $\lambda_{ex}$ = 468 nm and $\lambda_{em}$ = 500-700 nm, RBITC: $\lambda_{ex}$ 561 nm and $\lambda_{em}$ 570-700 nm; PtOEP: $\lambda_{ex}$ =542nm and $\lambda_{em}$ =570-700 nm and Alexa647: $\lambda_{ex}$ =650nm and $\lambda_{em}$ =670-700 nm. Values expressed as mean ± standard error (±SE) of five independent experiments (n=5).

### 3.4 Oxygen measurements are not affected by pH, Na$^+$, K$^+$ and H$_2$O$_2$ variations

Selectivity of oxygen sensors to the sole PO$_2$ is crucial for measuring correct values of oxygen levels, independently from other biological analytes in the surrounding microenvironment. In biological systems, pH is a parameter that can vary between tissues and cell compartments.[57] In addition, cations like Na$^+$ and K$^+$ play an important role in the regulation of cellular electrolyte metabolism, electric signalling in cells, transport of essential nutrients and determination of the membrane potential.[58] To evaluate selectivity at different pH values and cation concentrations, sensors, kept at PO$_2$ of 19.6 %, were incubated in pH-adjusted L-15 media in the physiological range (from 5.0 to 8.0) or in buffers at know Na$^+$/K$^+$ concentration in the physiological range (from 0 to 160 mM) and their emission signals were recorded at the plate reader. As shown in **Figures 2e-f**, pHs and oxygen were not correlated for both sensing microparticles. Indeed, the ratio between each sensitive and reference dyes, remained constantly independent of the pH variations. However, it should be noticed that the ratio values for Ru(dpp)-RBITC sensors showed less fluctuation around the mean value compared to the PtOEP-Alexa647 ones. This could be ascribed not only to the metal-organic complex itself, but also to the reference indicators (RBITC and Alexa647). Indeed, Alexa647 stability decreases in buffer as stated by the producers.[59] This assumption was further confirmed by the results recorded in presence of increasing concentration of Na$^+$/K$^+$ (**Figures 2g-h**). Ru(dpp)-RBITC sensors showed random fluctuation around the mean value at increasing concentration of $Na^+/K^+$ proving their selectivity towards oxygen even in presence of high cations concentrations. PtOEP-Alexa647 oxygen dosages appeared slightly altered at $K^+$ concentrations higher than 100 mM, confirming that Alexa647 stability is influenced by the ionic strength of the buffer employed during the measurements. However, this small interference does not affect extracellular oxygen monitoring where the highest $K^+$ concentration reaches approximately 30 mM in pathological conditions.[60, 61] Taken together, the results suggest that Ru(dpp)-RBITC sensors can be employed either for intra- or extracellular measurements whereas PtOEP-Alexa647 sensors can be better suited for extracellular monitoring of oxygen in those biological systems, such as the tumour microenvironment, that are notoriously affected by pH fluctuations and $Na^+/K^+$ variations.[4] Dissolved oxygen results 10-1000 times in excess compared to reactive oxygen species[62] (ROS) nevertheless, the amount of ROS compared to dissolved oxygen was investigated to evaluate sensors affinity also towards reactive oxygen species. For instance, Ru(dpp)-RBITC and PtOEP-Alexa647 sensors response was tested in presence of L15 medium enriched with known relevant concentrations of $H_2O_2$ in the physio-pathological range (0, 5, 10, 20 30 mM) at $PO_2$ 19.6%. The results reported in **Figure 2 i-j** show random fluctuations of Ru(dpp)-RBITC and PtOEP-Alexa647 ratios around their mean values, attesting a nonspecific response of the sensors toward ROS species proving that oxygen measurements are not affected by ROS variations. This could probably attributed to the PDMS-OH shell with excellent chemical stability (high resistance to acid, base or oil) and tunable adhesion to silica surfaces[63] that embeds and protects effectively the oxygen-sensitive Ru(dpp)/PtOEP dyes.

### 3.5 Oxygen sensing microparticles do not affect cell viability

Biosensors play a crucial role in quickly and precisely detecting biological analytes within both physiological and pathological systems, including cancer. Their utility extends to assessing the efficacy of anticancer chemotherapy agents, offering valuable insights into treatment outcomes.[9] To explore the suitability of the oxygen sensing microparticles for practical applications, we evaluated their cytotoxicity across three distinct tumor cell lines and on normal cells as cancer cell viability is typically higher than that of normal cells. HeLa, AsPC-1, and MCF7 tumor cells or 3T3 normal cells were incubated with either Ru(dpp)-RBITC or PtOEP-Alexa647 sensors for 0, 24, and 48 hours, followed by assessment of cell viability using the MTT assay. Untreated cells served as the control group. As reported in **Figure 3**, no significant cytotoxic effect was observed in all treated cells in comparison to their respective controls. Specifically, cells incubated with the oxygen microsensors exhibited consistent viability levels across various incubation time points, with no statistically significant differences in terms of cytotoxicity observed between the treated and untreated groups.

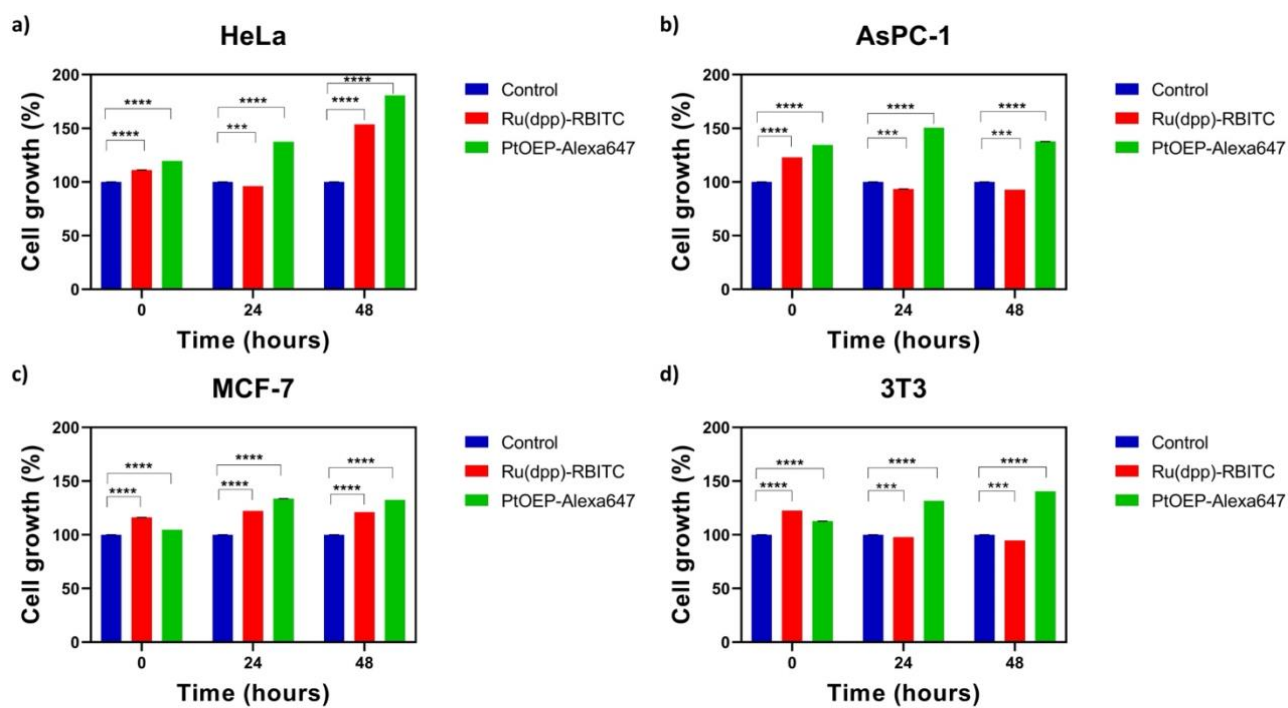

**Figure 3.** Biocompatibility of oxygen-sensitive microsensors on **a)** HeLa, **b)** AsPC-1, **c)** MCF-7 and **d)** 3T3 cell lines. Graphs indicating the absorbance of formazan crystals dissolved in acidified isopropanol after the incubation of tumor or normal cells with MTT reagent at 0, 24 and 48 hours of treatment with Ru(dpp)-RBITC or PtOEP-Alexa647 sensors. Values expressed as mean ± standard error (±SE) of three independent experiments (n=3). Statistical analysis: **** $p<0.0001$, *** $p<0.001$, Control *vs.* Ru(dpp)-RBITC or PtOEP-Alexa647.

The nontoxic behaviour of the microsensors could be mainly due to the bio-inert silica behaviour[64] and to the biocompatible oxygen permeable PDMS-OH sealer.[65] Furthermore, the micrometer size of the particles allows the reduction of surface energy which is the main cause of interaction of nanoparticles with cell organelles[66] and biomolecules.[67,68]

*3.6 Response of Ru(dpp)-RBITC and PtOEP-Alexa647 sensors to oxygen variations*

**Figure 4a** shows representative CLSM micrographs of Ru(dpp)-RBITC sensors at $PO_2$ of 1.3 %, 6.0 %, 11.7 % and 17.2 %, respectively. In **Figure 4b**, the intensity ratio $I_{Ru(dpp)}/I_{RBITC}$, obtained from CLSM images analyzed through image segmentation algorithms, is reported against the $PO_2$. In accordance with the physical properties of the oxygen-sensitive probe Ru(dpp),[37] the emission of the microsensors was highly dependent on the local dissolved oxygen. In particular, at low concentration of oxygen the Ru(dpp) fluorescence intensity (in green, false color) is high, while at high concentration of dissolved oxygen the Ru(dpp) fluorescence is quenched. Instead, as expected, the intensity of the reference dye, RBITC (red), does not change as dramatically as Ru(dpp) at high and low concentration of dissolved oxygen (**Figure 4a**). As represented in the calibration curve (fit parameters illustrated in the graph) of **Figure 4b**, the relationship between the fluorescence intensity ratio and dissolved oxygen is inversely proportional in strongly agreement with the Stern-Voltmer equation.[56]

Similarly, **Figure 4c-d** show the results obtained for the PtOEP-Alexa647 sensors. In **Figure 4c** representative confocal micrographs at $PO_2$ of 1.4 %, 6.1 %, 11.6 % and 17.1 % are reported and in **Figure 4d** the intensity ratios ($I_{PtOEP}/I_{Alexa647}$) recorded by the CLSM analysis are shown. Similarly to the Ru(dpp)-RBITC sensors, the PtOEP-Alexa647 oxygen sensors showed a behaviour in accordance with the photophysical properties of the oxygen sensing PtOEP dye.[35] Also here, the curve trend attests an inversely proportional relationship between PtOEP fluorescence and oxygen concentration in accordance with the Stern-Voltmer equation.[56] Importantly, these CLSM analyses are in accordance with spectrofluorimetric data and confirmed Ru(dpp)-RBITC sensors to be more sensitive and accurate in measuring the oxygen content than PtOEP-Alexa647. Indeed, Ru(dpp)-RBITC equation of the fit calibration curve has a higher slope (0.030 for Ru(dpp) and 0.020 for PtOEP) than that of PtOEP-Alexa647 sensors.

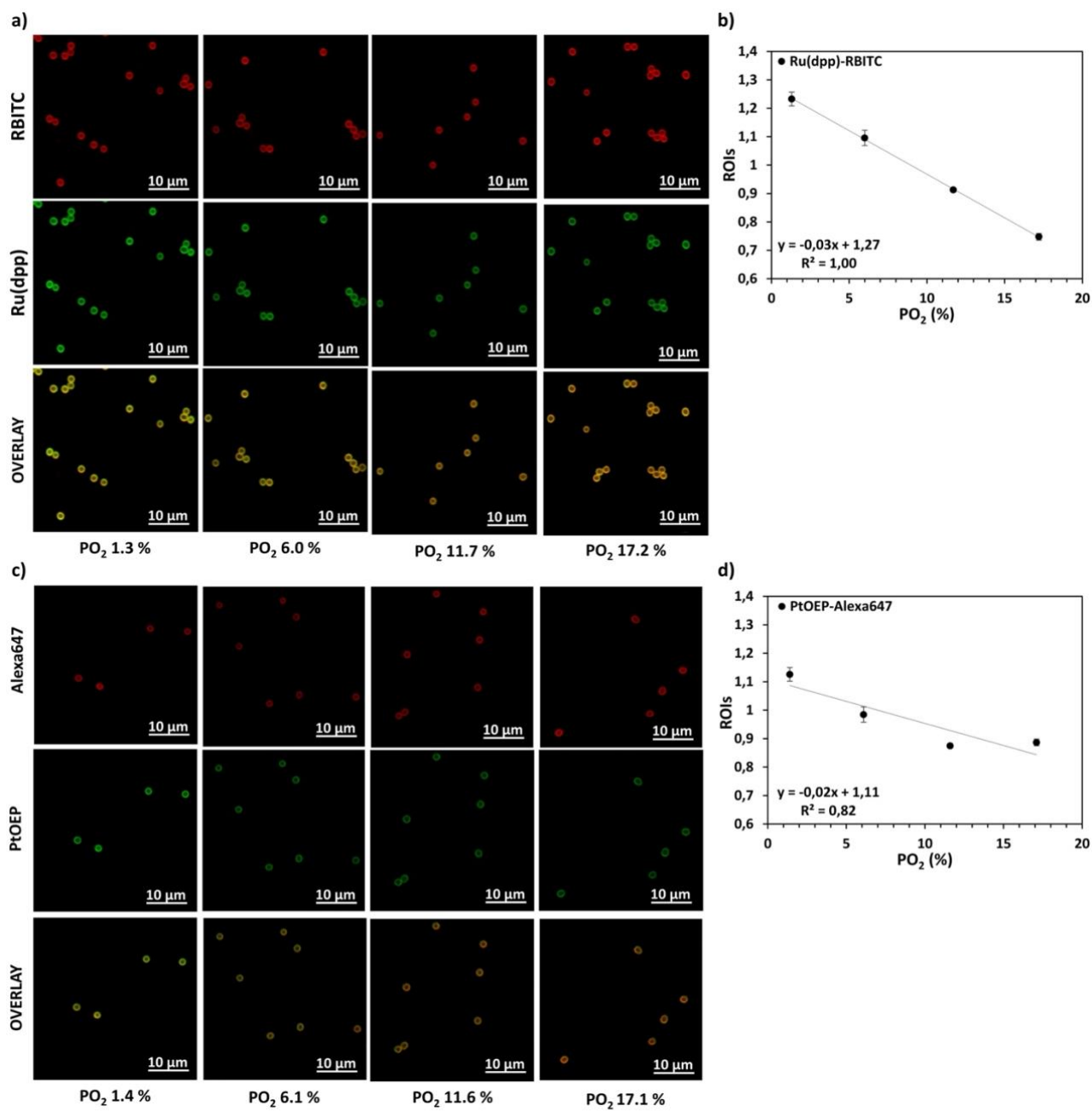

**Figure 4**. Oxygen-sensitive microsensors CLSM Calibration: **a)** Representative CLSM images showing the oxygen dependence of Ru(dpp)-RBITC's fluorescence. Green channel (Ru(dpp): $\lambda_{ex}$ 488 nm, $\lambda_{em}$ 489-630 nm), red channel (RBITC: $\lambda_{ex}$ 555 nm, $\lambda_{em}$ 556-630 nm) and overlay of the fluorescence channels are reported. Scale bars: 10 µm. **b)** Ratiometric calibration curve showing the fluorescence intensity ratio of green and red channels derived from CSLM micrographs. Values expressed as mean ± standard error (±SE) of three independent experiments (n=3). **c)** Representative CLSM maximum z-projection images showing the oxygen dependence of PtOEP-Alexa647's fluorescence. False green channel (PtOEP: $\lambda_{ex}$ 555nm, $\lambda_{em}$ 500-700 nm) and red channel (Alexa647: $\lambda_{ex}$ 639nm, $\lambda_{em}$ 640-700 nm), and overlay of the fluorescence channels are reported. Scale bars: 10 μm. **d)** Ratiometric calibration curve showing the fluorescence intensity ratio of false green and red channels derived from CLSM micrographs. Values expressed as mean ± standard error (±SE) of three independent experiments (n=3).

Since organic fluorescent indicators are affected by photobleaching during long exposures at CLSM,[69] we carried out photobleaching tests to evaluate the possibility of change in emission ratio. Cycles of 20 acquisitions, in the same field of view, at low and high $PO_2$, were recorded applying the same settings employed during the CLSM calibration of the sensors. **Figure S7** reports green and red channels, as well as the ratio trends of Ru(dpp)-RBITC sensing microparticles at $PO_2$ of 1.3% (**a-b**) and 17.2% (**c-d**), respectively. Ru(dpp)-RBITC is still responsive after 20 acquisitions, and the sensors appear reliable and robust with negligible emission intensity fluctuations and absence of ratio trend at both the monitored $PO_2$. The results proved that both reference and sensitive dyes have similar quenching behaviour thus, the ratio was always conserved. Regarding the PtOEP-Alexa647 sensors, dye emission intensities exhibited small fluctuations at both 1.4% (**Figure S8a-b**) and 17.1% (**Figure S8c-d**) of $PO_2$. However, being these small and random fluctuations, they did not alter the intensity ratio at high $PO_2$. On the contrary, despite the use of very low laser powers, Alexa647 reference dye was more affected by photobleaching at low $PO_2$ and this caused a slight alteration of the ratio. This finding is in accordance with spectrofluorimetric results and confirms that Alexa647 photostability decreases in buffer as stated by the producers.[59] Overall, Ru(dpp)-RBITC sensors were more promising in terms of oxygen sensitivity and photobleaching. Therefore, Ru(dpp)-RBITC sensors were employed for next biological studies.

### *3.7 Oxygen-sensing properties of Ru(dpp)-RBITC microparticles in in vitro 3D alginate scaffolds*

Next, the response of Ru(dpp)-RBITC oxygen sensors was tested within 3D alginate hydrogel scaffolds. To this aim, the sensors were embedded in alginate hydrogels and exposed to different oxygen concentrations ($PO_2$ range 1.5-17.1%) for 40 minutes before being imaged by CLSM. As shown **Figure S9a**, according to the optical properties of the oxygen-sensitive probe (Ru(dpp)) and the reference dye (RBITC), the microsensors at $PO_2$ 1.5% emit a strong fluorescence signal from Ru(dpp) probe that becomes quenched when the local oxygen concentration approaches atmospheric values ($PO_2$ 17.1%), while the red emission from RBITC remains unchanged. The sensors display a linear response in the range of oxygen percentage between 1.5 % and 17.1 %, with a correlation coefficient $R^2$=0,962 (**Figure S9b**).

Notably, the confocal setting (laser power and gain) was optimized in order to overcome attenuation of the signal due to the presence of the alginate hydrogel. Thus, the fluorescence signal stability of Ru(dpp)-RBITC microparticles was tested. Cycles of 20 acquisitions were recorded on z-stacks of the same alginate hydrogel embedding Ru(dpp)-RBITC sensors in L-15 medium, at low and high oxygen percentage, adopting the settings employed during CLSM calibration. Graphs reported in **Figure S10** shows the fluorescence intensity of individual Ru(dpp) and RBITC channels and their ratio at $PO_2$ of 1.2% (**Figure S10 a-b**) and 17.1% (**Figure S10 c-d**), respectively. The obtained results demonstrate that the alginate matrix and the selected laser power do not affect sensor's performance, indeed the fluorescence signal remains stable over time with a similar trend to that of the free sensors in L-15 medium (**Figure S7**). Remarkably, the sensitivity of the core-shell Ru(dpp)-RBITC microsensors was preserved after the encapsulation steps into thr hydrogel matrixes. Afterwards, pancreatic cancer cells, AsPC-1 cell line, and pancreatic stromal cells, PSCs, were selected to develop hybrid cell-seeded oxygen sensing hydrogels that mimic the three dimentional tumor microenvironment of pancreatic ductal adenocarcinoma (PDAC). The validation of the $O_2$-sensing hybrid hydrogels in an *in vitro* tumor model is crucial given the presence of oxygen gradients with multiple hypoxic regions within the tumor microenvironment.[70] In this context we selected PDAC because it is among the tumors considered severely hypoxic, with a low oxygen content around 0.7 %.[44] To noninvasively monitor oxygen variations over time and space, $O_2$-sensing hybrid hydrogels were produced (see Material and Methods and Scheme in **Figure S2**). In particular, PSCs and AsPC-1 cells were encapsulated in alginate hydrogels containing Ru(dpp)-RBITC oxygen sensors, and oxygen variations were monitored for 12 hours by time lapse CSLM imaging. In **Figure 5a** and **Figure S11a**, representative CLSM images of the whole system are reported. The intensity ratio of each oxygen sensor was calculated over time and the oxygen values, extracted from the calibration curve (**Figure S9b**), were converted in color. For the whole oxygen sensing alginate hydrogel, a 3D map was created for each time point (from 0 to 12 hours) for monitoring the dynamic variations of the oxygen sensors over time and space (**Figure 5b** and **Figure S11b**). In particular at the same time point (time 0 and time 12h) the sensors surrounding tumor and stromal cells in the alginate hydrogel show different colors at different positions along the axes. Moreover, a strong reduction in oxygen concentration is evident after 12 hours of monitoring as reported by the color change of the sensors from red to yellow in the colorimetric map. Notably, it is known that cancer cells response to reduced oxygenation is not always the same, since cell death or viability can occur depending on the time of exposure to hypoxia.[71] In fact, as reported in the **Figure S12,** fluctuations of oxygen levels were recorded during the time-lapse experiment reaching low level (PO$_2$ 2.7 ± 0.1 %) after 12 hours of monitoring. These data reveal both the presence of a heterogeneous gradient of oxygen in the extracellular environment and the good performance and reliability of the oxygen sensors in a 3D environment.

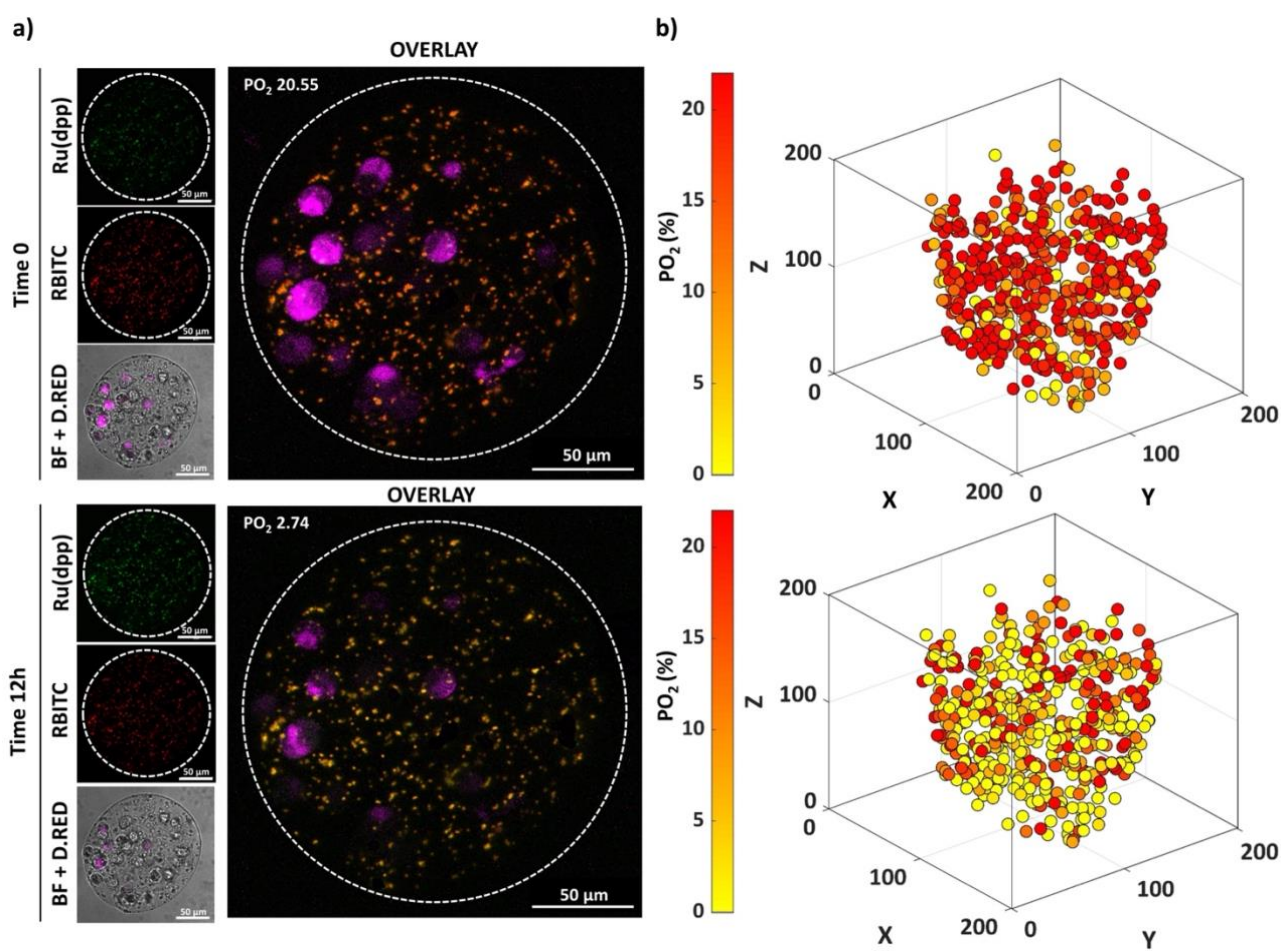

**Figure 5.** Extracellular PO$_2$ evaluation over time and space of 3D alginate hydrogels containing Ru(dpp)-RBITC oxygen sensors, stromal cells (PSC) and cancer cells (AsPC-1): **a)** Maximum projection of time-lapse images at time 0 and time 12 hours of Ru(dpp) (green channel), RBITC (red channel), PSC (bright field), AsPC-1 (deepRed, magenta) (field of view in 213.39 × 213.39 µm; step size 2.55 µm, z-projection of 29 sections); **b)** 3D scatter plots of Ru(dpp)-RBITC oxygen sensors of the experiment shown in **a** with relative oxygen colormaps.

## 4.0 CONCLUSIONS

Fluorescence-based sensing tools have emerged as a promising technology for detecting key metabolic parameters, such as oxygen, in biological systems, both *in vitro* and *in vivo*. However, their application in the field of 3D tumor models demand more precise methods to monitor analyte concentrations with high spatial and temporal resolution. In this study, we employed Tris(4,7-diphenyl-1,10-phenanthroline)ruthenium(II) dichloride (Ru(dpp)) and Platinum octaethylporphyrin (PtOEP) as oxygen sensing indicators for synthesizing two classes of silica core-double shell microparticle-based ratiometric sensors. We detailed the synthetic processes and properties of the resulting optical sensors, including morphology, performance, and toxicity assessment. Notably, the fabrication of these microsensors yielded simple, cost-effective, and innovative silica core-shell systems. Their response to oxygen variations was rapid, highly reproducible, and stable over time. Furthermore, cytotoxicity assays demonstrated the biocompatibility of the microsensors, making them suitable for *in vitro* cell studies. The enhanced oxygen-sensing microparticles, based on Ru(dpp)-RBITC dyes, were effectively utilized within a 3D *in vitro* construct mimicking the pancreatic tumor microenvironment. These sensors accurately tracked dynamic oxygen variations accompanying cellular proliferation, showcasing their potential for advanced biological studies. The present study highlights the potential of ratiometric core-double shell oxygen sensors in capturing spatio-temporal oxygen gradients within complex three-dimensional cellular architectures.

Temperature is a parameter that could affect the sensitivity of the oxygen sensing dyes[27, 55] however, these studies are intended to be done at 37°C thus, this issue will never be faced. Photobleaching over time is another potential limitation which can be observed if the sensors are used for prolonged periods. This is a general limitation that can be avoided using low intensity excitation lights and low exposure times. Moreover, Ru(dpp)-RBITC and PtOEP-Alexa647 are also compatible for PLIM analysis and are active to multiphoton imaging[24] that allow to reduce photobleaching issues and extensive sensors usage. Our findings indicate that these hybrid core-shell micro-silica sensors can be seamlessly integrated into 3D scaffolds resembling the extracellular matrix, offering a versatile and promising tool for conducting precise investigations into the role of hypoxia in cancer development and progression. This versatility places these sensors as valuable assets for delineating oxygen distribution both *in vitro* and *in vivo*. Moreover, their robustness, adaptability, and customizable nature enable their application to extend beyond cancer research. These sensors can find utility in environmental monitoring of oxygen levels in water samples or in tracking reactive oxygen species,[33] employing different combinations of dyes tailored to specific analysis setups.

**SUPPORTING INFORMATION:** Additional experimental graphs, details, methods, including photographs and experimental setup sketches.

# CONFLICTS OF INTEREST

There are no conflicts to declare.

# ACKNOWLEDGEMENTS

The authors acknowledge funding from the European Research Council (ERC) under the European Union's Horizon 2020 research and innovation program ERC Starting Grant "INTERCELLMED" (n. 759959), the European Union's Horizon 2020 research and innovation programme under grant agreement No. 953121 (FLAMIN-GO), the Associazione Italiana per la Ricerca contro il Cancro (AIRC) (MFAG-2019, n. 22902), the "Tecnopolo per la medicina di precisione" (TecnoMed Puglia) - Regione Puglia, n. B84I18000540002), the Italian Ministry of Research (MUR) in the framework of the National Recovery and Resilience Plan (NRRP), "NFFA-DI" Grant (n. B53C22004310006), "I-PHOQS" Grant (n. B53C22001750006) and under the complementary actions to the NRRP, "Fit4MedRob" Grant (PNC0000007, n. B53C22006960001), funded by NextGenerationEU and the PRIN 2022 (2022CRFNCP_PE11_PRIN2022).

# 5.0 REFERENCES

1. Ryter, S. W.; Choi, A. M., Heme oxygenase-1/carbon monoxide: from metabolism to molecular therapy. *American journal of respiratory cell and molecular biology* **2009,** *41* (3), 251-260.

2. Greer, S. N.; Metcalf, J. L.; Wang, Y.; Ohh, M., The updated biology of hypoxia-inducible factor. *The EMBO journal* **2012,** *31* (11), 2448-2460.

3. Eales, K. L.; Hollinshead, K. E.; Tennant, D. A., Hypoxia and metabolic adaptation of cancer cells. *Oncogenesis* **2016,** *5* (1), e190-e190.

4. Grasso, G.; Colella, F.; Forciniti, S.; Onesto, V.; Iuele, H.; Siciliano, A. C.; Carnevali, F.; Chandra, A.; Gigli, G.; del Mercato, L. L., Fluorescent nano-and microparticles for sensing cellular microenvironment: past, present and future applications. *Nanoscale Advances* **2023**.

5. Martínez-Reyes, I.; Chandel, N. S., Cancer metabolism: looking forward. *Nature Reviews Cancer* **2021,** *21* (10), 669-680.

6. Amao, Y., Probes and polymers for optical sensing of oxygen. *Microchimica Acta* **2003,** *143*, 1-12.

7. Van der Windt, G. J.; Chang, C. H.; Pearce, E. L., Measuring bioenergetics in T cells using a seahorse extracellular flux analyzer. *Current protocols in immunology* **2016,** *113* (1), 3.16 B. 1-3.16 B. 14.

8. Garedew, A.; Hütter, E.; Haffner, B.; Gradl, P.; Gradl, L.; Jansen-Dürr, P.; Gnaiger, E. In *High-resolution respirometry for the study of mitochondrial function in health and disease. The OROBOROS Oxygraph-2k*, Proceedings of the 11th Congress of the European Shock Society, Vienna, Austria (H Redl, ed) Bologna, Italy: Medimond International Proceedings, 2005; pp 107-111.

9. Hashem, M.; Weiler-Sagie, M.; Kuppusamy, P.; Neufeld, G.; Neeman, M.; Blank, A., Electron spin resonance microscopic imaging of oxygen concentration in cancer spheroids. *Journal of Magnetic Resonance* **2015,** *256*, 77-85.

10. Wolfbeis, O. S., Luminescent sensing and imaging of oxygen: Fierce competition to the Clark electrode. *BioEssays* **2015,** *37* (8), 921-928.

11. Gooz, M.; Maldonado, E. N., Fluorescence microscopy imaging of mitochondrial metabolism in cancer cells. *Frontiers in Oncology* **2023,** *13*, 1152553.

12. Prasad, S.; Chandra, A.; Cavo, M.; Parasido, E.; Fricke, S.; Lee, Y.; D'Amone, E.; Gigli, G.; Albanese, C.; Rodriguez, O., Optical and magnetic resonance imaging approaches for investigating the tumour microenvironment: state-of-the-art review and future trends. *Nanotechnology* **2020,** *32* (6), 062001.

13. Chandra, A.; Prasad, S.; Gigli, G.; del Mercato, L. L., Fluorescent nanoparticles for sensing. In *Frontiers of nanoscience*, Elsevier: 2020; Vol. 16, pp 117-149.

14. Yoshihara, T.; Hirakawa, Y.; Hosaka, M.; Nangaku, M.; Tobita, S., Oxygen imaging of living cells and tissues using luminescent molecular probes. *Journal of Photochemistry and Photobiology C: Photochemistry Reviews* **2017**, *30*, 71-95.

15. Periasamy, A.; Mazumder, N.; Sun, Y.; Christopher, K. G.; Day, R. N., FRET microscopy: basics, issues and advantages of FLIM-FRET imaging. *Advanced time-correlated single photon counting applications* **2015**, 249-276.

16. van Munster, E. B.; Gadella, T. W., Fluorescence lifetime imaging microscopy (FLIM). *Microscopy Techniques: -/-* **2005**, 143-175.

17. Becker, W., Fluorescence lifetime imaging–techniques and applications. *Journal of microscopy* **2012,** *247* (2), 119-136.

18. Wang, X. F.; Periasamy, A.; Herman, B.; Coleman, D. M., Fluorescence lifetime imaging microscopy (FLIM): instrumentation and applications. *Critical Reviews in Analytical Chemistry* **1992,** *23* (5), 369-395.

19. Rizzo, R.; Onesto, V.; Forciniti, S.; Chandra, A.; Prasad, S.; Iuele, H.; Colella, F.; Gigli, G.; Del Mercato, L. L., A pH-sensor scaffold for mapping spatiotemporal gradients in three-dimensional in vitro tumour models. *Biosensors and Bioelectronics* **2022,** *212*, 114401.

20. da Silva, A. d. S.; Dos Santos, J. H. Z., Stöber method and its nuances over the years. *Advances in Colloid and Interface Science* **2023**, 102888.

21. Chandra, A.; Prasad, S.; Iuele, H.; Colella, F.; Rizzo, R.; D'Amone, E.; Gigli, G.; Del Mercato, L. L., Highly sensitive fluorescent ph microsensors based on the ratiometric dye pyranine immobilized on silica microparticles. *Chemistry–A European Journal* **2021,** *27* (53), 13318-13324.

22. Onesto, V.; Forciniti, S.; Alemanno, F.; Narayanankutty, K.; Chandra, A.; Prasad, S.; Azzariti, A.; Gigli, G.; Barra, A.; De Martino, A., Probing single-cell fermentation fluxes and exchange networks via pH-sensing hybrid nanofibers. *ACS nano* **2022,** *17* (4), 3313-3323.

23. Moldero, I. L.; Chandra, A.; Cavo, M.; Mota, C.; Kapsokalyvas, D.; Gigli, G.; Moroni, L.; Del Mercato, L. L., Probing the pH Microenvironment of Mesenchymal Stromal Cell Cultures on Additive-Manufactured Scaffolds. *Small* **2020,** *16* (34), 2002258.

24. Chen, Y.; Guan, R.; Zhang, C.; Huang, J.; Ji, L.; Chao, H., Two-photon luminescent metal complexes for bioimaging and cancer phototherapy. *Coordination Chemistry Reviews* **2016,** *310*, 16-40.

25. Baggaley, E.; Gill, M. R.; Green, N. H.; Turton, D.; Sazanovich, I. V.; Botchway, S. W.; Smythe, C.; Haycock, J. W.; Weinstein, J. A.; Thomas, J. A., Dinuclear Ruthenium (II) Complexes as Two-Photon, Time-Resolved Emission Microscopy Probes for Cellular DNA. *Angewandte Chemie International Edition* **2014,** *53* (13), 3367-3371.

26. Baggaley, E.; Sazanovich, I. V.; Williams, J. G.; Haycock, J. W.; Botchway, S. W.; Weinstein, J. A., Two-photon phosphorescence lifetime imaging of cells and tissues using a long-lived cyclometallated N pyridyl^ C phenyl^ N pyridyl Pt (II) complex. *RSC Adv.* **2014,** *4* (66), 35003-35008.

27. Bansal, A.-K.; Holzer, W.; Penzkofer, A.; Tsuboi, T., Absorption and emission spectroscopic characterization of platinum-octaethyl-porphyrin (PtOEP). *Chemical physics* **2006,** *330* (1-2), 118-129.

28. Wang, X.-d.; Chen, H.-x.; Zhao, Y.; Chen, X.; Wang, X.-r., Optical oxygen sensors move towards colorimetric determination. *TrAC Trends in Analytical Chemistry* **2010,** *29* (4), 319-338.

29. Isbell, S.; Mendonsa, A.; Cash, K. J. In *Development of calcium and oxygen nanosensors for in-vivo diagnostics*, 2023 Spring Undergraduate Research Symposium, Colorado School of Mines. Arthur Lakes Library: 2023.

30. Sun, K.; Tang, Y.; Li, Q.; Yin, S.; Qin, W.; Yu, J.; Chiu, D. T.; Liu, Y.; Yuan, Z.; Zhang, X.; Wu, C., In Vivo Dynamic Monitoring of Small Molecules with Implantable Polymer-Dot Transducer. *ACS Nano* **2016,** *10* (7), 6769-6781.

31. Huo, S.; Carroll, J.; Vezzu, D. A., Design, synthesis, and applications of highly phosphorescent cyclometalated platinum complexes. *Asian Journal of Organic Chemistry* **2015,** *4* (11), 1210-1245.

32. Xu, R.; Wang, Y.; Duan, X.; Lu, K.; Micheroni, D.; Hu, A.; Lin, W., Nanoscale metal–organic frameworks for ratiometric oxygen sensing in live cells. *Journal of the American Chemical Society* **2016,** *138* (7), 2158-2161.

33. Li, W.; Jiang, G.-B.; Yao, J.-H.; Wang, X.-Z.; Wang, J.; Han, B.-J.; Xie, Y.-Y.; Lin, G.-J.; Huang, H.-L.; Liu, Y.-J., Ruthenium (II) complexes: DNA-binding, cytotoxicity, apoptosis, cellular localization, cell cycle arrest, reactive oxygen species, mitochondrial membrane potential and western blot analysis. *Journal of photochemistry and photobiology B: Biology* **2014,** *140*, 94-104.

34. Lee, G.; Jun, Y.; Jang, H.; Yoon, J.; Lee, J.; Hong, M.; Chung, S.; Kim, D.-H.; Lee, S., Enhanced oxygen permeability in membrane-bottomed concave microwells for the formation of pancreatic islet spheroids. *Acta biomaterialia* **2018,** *65*, 185-196.

35. Zhang, K.; Zhang, H.; Li, W.; Tian, Y.; Li, S.; Zhao, J.; Li, Y., PtOEP/PS composite particles based on fluorescent sensor for dissolved oxygen detection. *Materials Letters* **2016,** *172*, 112-115.

36. Lemon, C. M.; Karnas, E.; Bawendi, M. G.; Nocera, D. G., Two-photon oxygen sensing with quantum dot-porphyrin conjugates. *Inorganic chemistry* **2013,** *52* (18), 10394-10406.

37. Bukowski, R. M.; Ciriminna, R.; Pagliaro, M.; Bright, F. V., High-performance quenchometric oxygen sensors based on fluorinated xerogels doped with [Ru (dpp) 3] 2+. *Analytical chemistry* **2005,** *77* (8), 2670-2672.

38. Jin, C.; Guan, R.; Wu, J.; Yuan, B.; Wang, L.; Huang, J.; Wang, H.; Ji, L.; Chao, H., Rational design of NIR-emitting iridium (III) complexes for multimodal phosphorescence imaging of mitochondria under two-photon excitation. *Chemical Communications* **2017,** *53* (75), 10374-10377.

39. Zhang, K. Y.; Gao, P.; Sun, G.; Zhang, T.; Li, X.; Liu, S.; Zhao, Q.; Lo, K. K.-W.; Huang, W., Dual-phosphorescent iridium (III) complexes extending oxygen sensing from hypoxia to hyperoxia. *Journal of the American Chemical Society* **2018,** *140* (25), 7827-7834.

40. Süss-Fink, G., Water-soluble arene ruthenium complexes: From serendipity to catalysis and drug design. *Journal of Organometallic Chemistry* **2014,** *751*, 2-19.

41. Shi, C.; Anson, F. C., A simple method for examining the electrochemistry of metalloporphyrins and other hydrophobic reactants in thin layers of organic solvents interposed between graphite electrodes and aqueous solutions. *Analytical chemistry* **1998,** *70* (15), 3114-3118.

42. Figueiredo, T. L.; Johnstone, R. A.; Sørensen, A. M. S.; Burget, D.; Jacques, P., Determination of Fluorescence Yields, Singlet Lifetimes and Singlet Oxygen Yields of Water-Insoluble Porphyrins and Metalloporphyrins in Organic Solvents and in Aqueous Media. *Photochemistry and photobiology* **1999,** *69* (5), 517-528.

43. Lee, L. C. C.; Lo, K. K. W., Strategic Design of Luminescent Rhenium (I), Ruthenium (II), and Iridium (III) Complexes as Activity-Based Probes for Bioimaging and Biosensing. *Chemistry–An Asian Journal* **2022,** *17* (22), e202200840.

44. Geyer, M.; Schreyer, D.; Gaul, L.-M.; Pfeffer, S.; Pilarsky, C.; Queiroz, K., A microfluidic-based PDAC organoid system reveals the impact of hypoxia in response to treatment. *Cell Death Discovery* **2023,** *9* (1), 1-8.

45. Delle Cave, D.; Rizzo, R.; Sainz Jr, B.; Gigli, G.; Del Mercato, L. L.; Lonardo, E., The revolutionary roads to study cell–cell interactions in 3d in vitro pancreatic cancer models. *Cancers* **2021,** *13* (4), 930.

46. Ojeda-Mendoza, G. J.; Contreras-Tello, H.; Rojas-Ochoa, L. F., Refractive index matching of large polydisperse silica spheres in aqueous suspensions. *Colloids and Surfaces A: Physicochemical and Engineering Aspects* **2018,** *538*, 320-326.

47. Schneider, C.; Rasband, W.; Eliceiri, K., NIH image to ImageJ: 25 years of image analysis. Nat Meth 9 (7): 671–675. 2012.

48. Rizzo, R.; Onesto, V.; Forciniti, S.; Chandra, A.; Prasad, S.; Iuele, H.; Colella, F.; Gigli, G.; Loretta, L., A pH-Sensor scaffold for mapping spatiotemporal gradients in three-dimensional in vitro tumour models. *Biosensors and Bioelectronics* **2022**, 114401.

49. Alemanno, F.; Cavo, M.; Delle Cave, D.; Fachechi, A.; Rizzo, R.; D'Amone, E.; Gigli, G.; Lonardo, E.; Barra, A.; Del Mercato, L. L., Quantifying heterogeneity to drug response in cancer–stroma kinetics. *Proceedings of the National Academy of Sciences* **2023,** *120* (11), e2122352120.

50. Rizzo, R.; Onesto, V.; Morello, G.; Iuele, H.; Scalera, F.; Forciniti, S.; Gigli, G.; Polini, A.; Gervaso, F.; del Mercato, L. L., pH-sensing hybrid hydrogels for non-invasive metabolism monitoring in tumor spheroids. *Materials Today Bio* **2023**, 100655.

51. Otsu, N., A threshold selection method from gray-level histograms. *IEEE transactions on systems, man, and cybernetics* **1979,** *9* (1), 62-66.

52. Kurdyukov, D. A.; Eurov, D. A.; Kirilenko, D. A.; Sokolov, V. V.; Golubev, V. G., Tailoring the size and microporosity of Stöber silica particles. *Microporous and Mesoporous Materials* **2018,** *258*, 205-210.

53. Wu, S.-H.; Mou, C.-Y.; Lin, H.-P., Synthesis of mesoporous silica nanoparticles. *Chemical Society Reviews* **2013,** *42* (9), 3862-3875.

54. Shin, M. S.; Kim, S. J.; Kim, I. Y.; Kim, N. G.; Song, C. G.; Kim, S. I., Swollen behavior of crosslinked network hydrogels based on poly (vinyl alcohol) and polydimethylsiloxane. *Journal of applied polymer science* **2002,** *85* (5), 957-964.

55. Kocincova, A. S.; Borisov, S. M.; Krause, C.; Wolfbeis, O. S., Fiber-optic microsensors for simultaneous sensing of oxygen and pH, and of oxygen and temperature. *Analytical chemistry* **2007,** *79* (22), 8486-8493.

56. Gehlen, M. H., The centenary of the Stern-Volmer equation of fluorescence quenching: From the single line plot to the SV quenching map. *Journal of Photochemistry and Photobiology C: Photochemistry Reviews* **2020,** *42*, 100338.

57. Grabe, M.; Oster, G., Regulation of organelle acidity. *The Journal of general physiology* **2001,** *117* (4), 329-344.

58. Zacchia, M.; Abategiovanni, M. L.; Stratigis, S.; Capasso, G., Potassium: from physiology to clinical implications. *Kidney Diseases* **2016,** *2* (2), 72-79.

59. Thermo Fisher Scientific Alexa Fluor 647 dye. https://www.thermofisher.com/it/en/home/life-science/cell-analysis/fluorophores/alexa-fluor-647.html#.

60. Eil, R.; Vodnala, S. K.; Clever, D.; Klebanoff, C. A.; Sukumar, M.; Pan, J. H.; Palmer, D. C.; Gros, A.; Yamamoto, T. N.; Patel, S. J., Ionic immune suppression within the tumour microenvironment limits T cell effector function. *Nature* **2016,** *537* (7621), 539-543.

61. Tan, J. W.; Folz, J.; Kopelman, R.; Wang, X., In vivo photoacoustic potassium imaging of the tumor microenvironment. *Biomedical Optics Express* **2020,** *11* (7), 3507-3522.

62. Murphy, M. P.; Bayir, H.; Belousov, V.; Chang, C. J.; Davies, K. J.; Davies, M. J.; Dick, T. P.; Finkel, T.; Forman, H. J.; Janssen-Heininger, Y., Guidelines for measuring reactive oxygen species and oxidative damage in cells and in vivo. *Nature metabolism* **2022,** *4* (6), 651-662.

63. Liu, J.; Yao, Y.; Li, X.; Zhang, Z., Fabrication of advanced polydimethylsiloxane-based functional materials: Bulk modifications and surface functionalizations. *Chemical Engineering Journal* **2021,** *408*, 127262.

64. Murugadoss, S.; Lison, D.; Godderis, L.; Van Den Brule, S.; Mast, J.; Brassinne, F.; Sebaihi, N.; Hoet, P. H., Toxicology of silica nanoparticles: an update. *Archives of toxicology* **2017,** *91* (9), 2967-3010.

65. Pergal, M. V.; Nestorov, J.; Tovilović, G.; Ostojić, S.; Gođevac, D.; Vasiljević-Radović, D.; Djonlagić, J., Structure and properties of thermoplastic polyurethanes based on poly (dimethylsiloxane): assessment of biocompatibility. *Journal of Biomedical Materials Research Part A* **2014,** *102* (11), 3951-3964.

66. Croissant, J. G.; Fatieiev, Y.; Khashab, N. M., Degradability and clearance of silicon, organosilica, silsesquioxane, silica mixed oxide, and mesoporous silica nanoparticles. *Advanced materials* **2017,** *29* (9), 1604634.

67. Vertegel, A. A.; Siegel, R. W.; Dordick, J. S., Silica nanoparticle size influences the structure and enzymatic activity of adsorbed lysozyme. *Langmuir* **2004,** *20* (16), 6800-6807.

68. Shang, W.; Nuffer, J. H.; Muñiz-Papandrea, V. A.; Colón, W.; Siegel, R. W.; Dordick, J. S., Cytochrome c on silica nanoparticles: influence of nanoparticle size on protein structure, stability, and activity. *Small* **2009,** *5* (4), 470-476.

69. Nwaneshiudu, A.; Kuschal, C.; Sakamoto, F. H.; Anderson, R. R.; Schwarzenberger, K.; Young, R. C., Introduction to confocal microscopy. *Journal of Investigative Dermatology* **2012,** *132* (12), 1-5.

70. Campillo, N.; Falcones, B.; Otero, J.; Colina, R.; Gozal, D.; Navajas, D.; Farre, R.; Almendros, I., Differential oxygenation in tumor microenvironment modulates macrophage and cancer cell crosstalk: novel experimental setting and proof of concept. *Frontiers in oncology* **2019,** *9*, 43.

71. Vaupel, P., The role of hypoxia-induced factors in tumor progression. *The oncologist* **2004,** *9* (S5), 10-17.